\documentclass[aps,prl,showpacs,twocolumn,floats]{revtex4}
\usepackage{graphicx}
\usepackage{amsmath}
\usepackage{psfrag}

\input{epsf}

\begin{document}

\title{Steady Stokes flow with long-range correlations,
fractal Fourier spectrum, \\and anomalous transport}
\author{Michael A. Zaks}

\affiliation{Institute of Physics, Humboldt University of Berlin,
10115 Berlin, Germany}
\author{Arthur V. Straube\footnote{Current e-mail: straube@stat.physik.uni-potsdam.de \\[0.5mm] {\mbox{} Paper published in Phys. Rev. Lett {\bf{89}},
244101 (2002)}}} \affiliation{Physics Department, Perm State
University, 614990 Perm, Russia}

\begin{abstract}
We consider viscous two-dimensional steady flows of incompressible
fluids past doubly periodic arrays of solid obstacles. In a class
of such flows, the autocorrelations for the Lagrangian observables
decay in accordance with the power law, and the Fourier spectrum
is neither discrete nor absolutely continuous. We demonstrate that
spreading of the droplet of tracers in such flows is anomalously
fast. Since the flow is equivalent to the integrable Hamiltonian
system with 1 degree of freedom, this provides an example of
integrable dynamics with long-range correlations, fractal power
spectrum, and anomalous transport properties.

\end{abstract}

\maketitle

The last two decades have witnessed growing interest in
theoretical and applied aspects of chaotic advection in steady or
temporally periodic velocity fields of hydrodynamical
flows~\cite{Aref-84,Ottino-89}. The contrast between the trivial
behavior of the field and the complicated dynamics of individual
fluid particle is due to the fact that the field is usually
described by means of Eulerian variables in the laboratory
reference frame, whereas for the fluid particle the Lagrangian
description is more appropriate, since its reference frame is
carried by the flow.
An Eulerian observer performs measurements
in the fixed point of the physical space and registers
a steady or  periodic process.
For a Lagrangian observer, physical space
turns into phase space, streamlines become phase trajectories,
a ``measuring device'' is advected along these trajectories
and explores different regions of the flow pattern.
Therefore  the motion of the fluid particle can be chaotic in spite
of the time-independent state at any given place.
A consequence of this dynamics is intensive stirring and mixing.
Examples of chaotic advection are widespread,
from microscopical flows  of nanotechnology and cellular biology
to large-scale phenomena in astrophysics and geophysics; for a list
of references see a recent review~\cite{Aref-2002}.

To exhibit Lagrangian chaos, a flow of incompressible fluid must
be three dimensional, time-dependent, or both. In two-dimensional
setup, there is no place for exponential divergence of
streamlines, and in conventional examples the motion of a typical
fluid particle is periodic and well correlated. Below, we discuss
a class of steady two-dimensional flows  for which the
correlations decay and power spectra are fractal.

Consider a slow steady flow of a viscous incompressible fluid past
the square lattice of parallel solid circular cylinders. We
restrict ourselves to the ``transverse'' case when the component
of the flow along the cylinder axes is absent. This geometry is
typical, e.g., for the cooling flows past the rods of the reactor;
on the other hand, it serves as the starting point for the
analysis of fluid motions in composite and porous media. Usual
assumptions of the creeping flow allow one to neglect nonlinear
terms in the Navier-Stokes equations and reduce them to the Stokes
equations for the velocity $\mathbf{V}$ and pressure $p$:
\begin{equation}
-\nabla p+\eta\Delta \mathbf{V}=0,\;\;\rm{div}\,\bbox{V}=0
\label{stokes_v}
\end{equation}
where $\eta$ is the viscosity of the fluid.
Boundary conditions enforce periodicity on the borders
of the elemental cell of the lattice, as well as vanishing velocity
on the  border $\partial\Gamma$ of the cylinder $\Gamma$.
If the axes $x$ and $y$ are aligned along the axes of the lattice,
and the cell size is chosen as the length unit, the boundary conditions are:
\begin{equation}
\mathbf{V}(0,y)=\mathbf{V}(1,y),\;\;\;
\mathbf{V}(x,0)=\mathbf{V}(x,1),\;\;\;
\mathbf{V}\bigl|_{\partial\Gamma}=0 \label{b_cond}
\end{equation}

Formulation of the boundary problem is completed by
prescribing the components of mean flow along both axes:
\begin{equation}
\int_0^1 V_x dy=\alpha,\;\int_0^1 V_y dx=\beta
\label{int_cond}
\end{equation}

Equations (\ref{stokes_v})-(\ref{int_cond})  were treated in terms
of periodic Green's function by H.~Hasimoto who evaluated the drag
on a cylinder~\cite{Hasimoto-59}; the fundamental solution was
expressed in terms of series containing elliptic
functions~\cite{Hasimoto-preprint}. Permeability of the array of
cylinders was estimated in~\cite{Sangani-Yao}.
As every two-dimensional flow of incompressible fluid, the problem
can be cast into a Hamiltonian frame by introducing the stream
function $\Psi(x,y)$: $V_x=\partial\Psi/\partial y$ and
$V_y=-\partial\Psi/\partial x$. Motions along the streamlines are
governed by the integrable Hamiltonian system with 1 degree of
freedom in which  $V_x$ and $V_y$ are canonical variables, and
$\Psi$ plays the role of energy. In terms of $\Psi$
Eq.~(\ref{stokes_v}) turns into the biharmonic equation $\Delta
\Delta \Psi=0$. No-slip boundary conditions require
$\partial\Psi/\partial x\bigl|_{\partial\Gamma}=
\partial\Psi/\partial y\bigl|_{\partial\Gamma}=0$,
whereas conditions for mean flow and periodicity are ensured by putting
$\Psi(x,y)=\alpha y -\beta x +\Phi(x,y)$ where $\Phi$ has
period 1 in both arguments.

\begin{figure}[h]
\centerline{\epsfxsize=0.48\textwidth\epsfbox{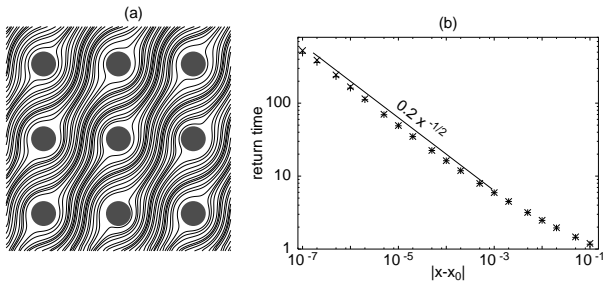}}\vspace{-3mm}
\caption{(a) Flow pattern with $\alpha/\beta=(\sqrt{5}-1)/2$; (b)
divergence of passage time for streamlines near the cylinder.
Crosses: numerical values, straight line: power law $\sim
|x-x_0|^{-1/2}$ } \label{pattern}
\end{figure}

We solved the biharmonic equation numerically, imposing periodic
boundary conditions upon a polar grid with 600x1200 nodes; for
tracer dynamics, the 2nd order interpolation scheme was used.
Figure~\ref{pattern}(a) presents the flow pattern with inclination
$\alpha/\beta=(\sqrt{5}-1)/2$ and radius of the cylinder $1/6$.
\\[-3mm]

Since the flow is steady, in every point of the laboratory
reference frame the values of velocity and pressure are time
independent. The picture looks different in a frame attached to a
neutral tracer carried by the flow. We characterize it in terms of
normalized autocorrelation, defined a time-dependent observable
$\xi(t)$ as
\begin{equation}
C(\tau)=\frac{\langle \xi(t)\xi(t+\tau)\rangle -\langle \xi(t)\rangle^2}
{\langle \xi^2(t)\rangle -\langle \xi(t)\rangle^2}
\label{cor_def}
\end{equation}
where averaging over $t$ is performed. Since autocorrelation is
the Fourier transform  of the power spectrum (the Wiener-Khinchin
theorem), its properties are related to the nature of the Fourier
spectral measure. For periodic and quasiperiodic dynamics,
$C(\tau)$ displays series of peaks with unit or asymptotically
unit amplitude at arbitrarily large values of $\tau$; this
corresponds to the discrete (pure-point) power spectrum.  In case
of chaotic behavior (impossible in two-dimensional dynamics)
autocorrelation decays exponentially, and the power spectrum is
absolutely continuous with respect to the Lebesgue measure.

\begin{figure}[b]
\centerline{\epsfxsize=0.44\textwidth\epsfbox{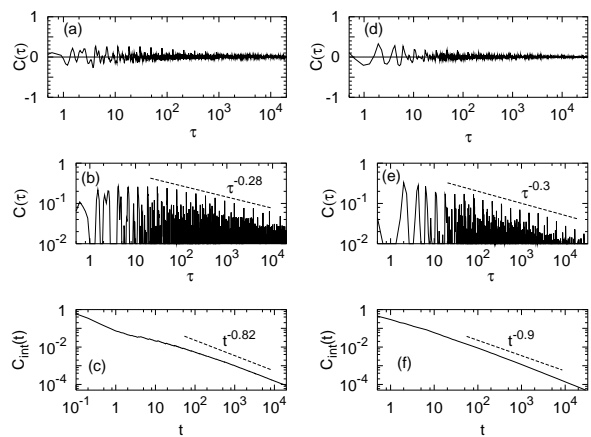}}\vspace{-3mm}
\caption{Autocorrelation and integrated autocorrelation. (a)-(c):
Stokes flow (\ref{stokes_v})-(\ref{int_cond}); (d)-(f): special
flow (\ref{specflow}).
Dashed lines: power laws.}
\label{corfun}
\end{figure}

Qualitative features of the autocorrelation are independent of the
choice of a generic observable. We take the absolute value
$|\bbox{V}|$ of the velocity and measure it for the fluid particle
moving along the streamline of the flow. The plot of $C(\tau)$ is
presented in Fig.~\ref{corfun}(a). The largest plotted values of
$\tau$ correspond to the passage through, on the average, $3\times
10^4$ lattice cells in the $y$ direction. We observe that
autocorrelation decays, in spite of the ordered laminar structure
of the flow. The decay is slower than exponential; as seen in
Fig.~\ref{corfun}(b), the highest peaks of $C(\tau)$ which form a
log-periodic lattice, decrease in accordance with the  power law
$C(\tau)\sim \tau^{-0.28}$. Power-law decay of autocorrelation
(``long-range correlations'') is abundant in physics of critical
and nearly critical states~\cite{Stanley-93} and has been
recovered in many natural phenomena, from DNA
sequences~\cite{Li-92,Arneodo-95,Audit-2001} to atmospheric
variability~\cite{meteo-98}. However, to our knowledge, such decay
has been reported neither in the context of steady two-dimensional
flows, nor in a more general context of integrable autonomous
Hamiltonian systems with 1 degree of freedom, where the
correlations are usually not supposed to decay at all. Our results
show that integrability does not guarantee against the decay of
correlations.
Properties of autocorrelation characterize both qualitatively and
quantitatively the set which supports the Fourier spectral
measure. A useful tool is the integrated autocorrelation $C_{\rm
int}(t)=t^{-1}\int_0^t C(\tau)^2 d\tau$: if $C_{\rm int}(t)$
vanishes in the limit  $t\to\infty$, power spectrum does not
include the discrete component. Furthermore, $C_{\rm int}(t)\sim
t^{-D_2}$ where $D_2$ is the correlation dimension of the spectral
measure~\cite{Ketzmerick}. The decay of $C_{\rm int}(t)$ over
several orders of magnitude in Fig.~\ref{corfun}(c) indicates the
absence of the discrete component in the Fourier spectrum of the
tracer velocity. Estimation of the curve slope yields $D_2\approx
0.82$; hence the set which supports spectral measure, is fractal.
Apparently, in this intermediate situation between chaos and order
the Fourier spectrum of velocity is singular continuous: neither
discrete, nor absolutely continuous.

Origin of this unconventional state can be understood if the
viscous flow is viewed as a dynamical system. From this point of
view,  Eqs~(\ref{stokes_v})-(\ref{int_cond}) describe the
area-preserving vector field $\dot{x}=V_x(x,y),\;\dot{y}=V_y(x,y)$
on a 2-torus whose rotation number $\sigma$ equals $\alpha/\beta$.
Peculiarity of this field is its identical vanishing on the
cylinder border $\partial\Gamma$: no-slip boundary conditions turn
each point of $\partial\Gamma$ into the fixed point. Below, we
deal with irrational values of $\sigma$. Most of the existing work
on dynamics on 2-tori, starting with the classical paper
in~\cite{Kolmogorov}, refers to fields without fixed points.
Consider a mapping of a line transversal to the flow (e.g. the
boundary $y=0$ of the square) onto itself. In absence of fixed
points the ``return time'' $\tau_{\rm ret}(x)$ needed for one
iteration of this mapping (a turn around the torus) is bounded. As
a consequence,  dynamics of the flow is adequately reproduced by
the circle map $x\to (x+\sigma)\;{\rm mod}\, 1$. For a passive
tracer in such quasiperiodic flow, the Fourier spectrum is
discrete, and the autocorrelation does not decay.

Presence of stagnation points in which the velocity vanishes
qualitatively changes dynamics. In this situation  $\tau_{\rm
ret}(x)$ as a function of $x$ diverges, and equivalence with the
Poincar\'e map is invalidated: the flow with continuous power
spectrum can induce the map with purely discrete
spectrum~\cite{scsds-95}. Near the streamline which leads into a
generic isolated stagnation point, the return time has a
logarithmic singularity. Such stagnation points exist in forced
viscous flows with doubly periodic arrays of steady vortices, and
the Fourier spectra in these flows are
fractal~\cite{Zaks-Pikovsky-Kurths-96}. Logarithmic singularities
of return times are too weak to make the flow
mixing~\cite{Kochergin-76}. Stronger, power-like singularities can
produce mixing ~\cite{Kochergin-75}; however, for isolated
stagnation points such singularities occur only in degenerate,
structurally unstable cases.

In the flow pattern of (\ref{stokes_v})-(\ref{int_cond}) the whole
curve $\partial\Gamma$ can be viewed as a structurally stable
family of stagnation points. Fluid particles exhibit the strong
slowdown during their motion along the edge of the obstacle. The
collective effect of this continuum of equilibria manifests itself
in the power-law divergence of return time: let $(x_0,0)$ lie on
the streamline which separates the particle paths passing the
obstacle ``on the left'' from the paths passing ``on the right,''
then $\tau_{\rm ret}(x)\sim 1/\sqrt{|x-x_0|}$
[Fig.~\ref{pattern}b]. The existence of this singularity ensures
the decay of correlations and creates continuous component in the
power spectrum.\\[-5mm]

\begin{figure}[h]
\centerline{\epsfxsize=0.48\textwidth\epsfbox{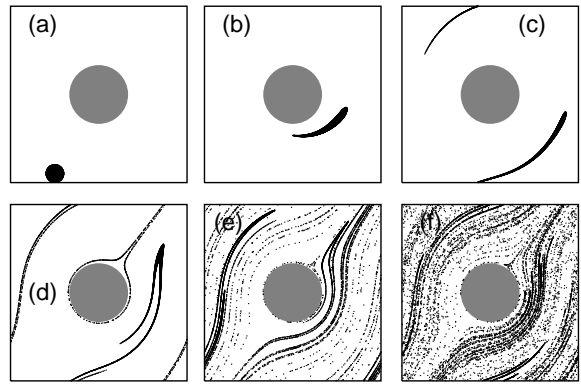}}\vspace{-3mm}
\caption{Cloud of passive tracers carried by the flow.\\
(a) $t$=0; (b) $t$=0.5; (c) $t$=2; (d) $t$=5; (e) $t$=20; (f) $t$=100. }
\label{droplet}
\end{figure}

Let us proceed from advection of individual passive tracer to the
transport of the ensemble of tracers by the flow
(\ref{stokes_v})-(\ref{int_cond}). Figure~\ref{droplet} shows the
temporal evolution of initially compact droplet formed by $10^4$
fluid particles. Nonisochronicity of rotations around the torus
stretches the droplet. The strongest effect is observed during the
passages in which the obstacle is sandwiched between the
streamlines: some particles hover in the vicinity of the obstacle
for the very long time, whereas the particles which pass at a
larger distance, are long gone. Because of irrational rotation
number, each streamline is dense on the torus; hence, each
particle passes arbitrarily close to the cylinder where its
velocity becomes arbitrarily small. As a result, after a hundred
of  passages the particles are spread over virtually the whole
area of the square. Naturally, this mixing is slower and less
efficient than mixing enabled by chaotic advection in the
situation of Lagrangian chaos~\cite{Aref-84,Ottino-89}.
Nevertheless, ultimately it leads to the same result.

\begin{figure}[b]
\centerline{\epsfxsize=0.3\textwidth\epsfbox{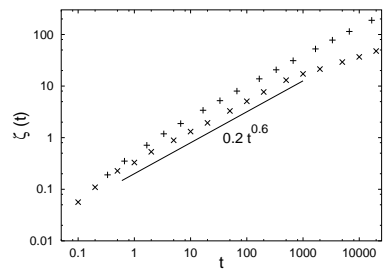}}\vspace{-3mm}
\caption{Growth of variance in the ensemble of particles
transported by the flow. Crosses: Stokes flow
(\ref{stokes_v})-(\ref{int_cond}); pluses: special flow
(\ref{specflow}) with $C=0.3$; straight line: power law $t^{0.6}$.
} \label{transport}
\end{figure}

To quantify the rate of spreading of the droplet on the infinite
plane, we introduce the displacement $\xi_y(t)\equiv y(t)-y(0)$
and estimate the time growth of the variance
$\zeta^2(t)=\langle\xi_y^2(t)\rangle-\langle\xi_y(t)\rangle^2$  by
averaging over all initial positions within the unit cell.
Figure~\ref{transport} shows that growth of $\zeta(t)$ is
described by the power law whose exponent exceeds the ``normal
diffusion'' value $1/2$.

For larger values of $t$ the computed values of $\zeta(t)$
undershoot the expected power law. We assign this discrepancy to
insufficient statistics: slow saturation in the dependence of
$\zeta$ on the length of the sample over which the averaging is
made. As $t$ grows, the main contribution into $\zeta(t)$ is made
by rare ``large events'': passages extremely close to the
obstacle. Apparently, the longest numerically affordable to us
trajectories with $3\times 10^6$ turns around the torus ($t\approx
2.8\times 10^6$) are still insufficient. In order to extend the
range of $t$ in which the estimates of $\zeta(t)$ are reliable, we
have used the model which is computationally much less expensive:
the ``special flow'' (flow over the mapping), often employed in
ergodic theory~\cite{Cornfeld-Fomin-Sinai-82}. For the map $u\to
f(u)$ and the function $\tau(u)$, this dynamical system is
introduced on the part of the plane $u,v$ defined by $0\leq v <
\tau(u)$, as follows: the orbit starts at $t=0$ in the point
$(u_0,v_0)$, and moves with unit velocity parallel to $v$-axis
until reaching the border $\tau(u_0)$; from there it
instantaneously jumps to the point $(f(u),0)$, begins the next
stage of motion along $v$, and so on. Obviously, on the $u$ axis
$f(u)$ is the Poincar\'e mapping, and $\tau(u)$ is the return
time. The trajectory is described by the expressions \\[-4mm]
\begin{equation}
u_t=f^n(u_0),\;
v_t=v_0+t-\sum_{j=0}^{n-1}\tau\bigl(f^j(u_0)\bigr)\\[-2mm]
\label{specflow}
\end{equation}
with integer $n=n(t)$ uniquely defined by inequalities \\[-4mm]
\begin{equation}
\sum_{j=0}^{n-1}\tau\bigl(f^j(u_0)\bigr) < v_0+t <
\sum_{j=0}^n\tau\bigl(f^j(u_0)\bigr)
\label{inequal}
\end{equation}

The special flow for our hydrodynamical problem,
combines the circle map $f(u)=(u+\alpha/\beta)\,{\rm mod}\,1$
with the return time $\tau(u)=C/\sqrt{|u-1/2|}$.
Properties of autocorrelation for the flow (\ref{specflow}),
presented in Figs.~\ref{corfun}(d)-\ref{corfun}(f), are in
qualitative and good quantitative accordance with the respective
characteristics of the Stokes flow: autocorrelation decays as the
power law, and the Fourier spectrum  is fractal. To evaluate the
transport in (\ref{specflow}) we used the lift of  $f(u)$ onto the
real axis. Estimates obtained from $10^{10}$ iterations of $f$ are
shown in Fig.~\ref{transport} by pluses. Again, we see the
anomalously fast growth of the variance.

These unusual transport properties are related to the distribution
of return times.  For $T$ large, the proportion of returns with
duration between $T$ and $T+dt$ decreases as $\sim T^{-3}dt$; the
second moment of this distribution (mean square of the return
time) is infinite. This is reminiscent of  L\'evy flights: random
walks with power-law statistics and infinite moments. Widespread
in physics, L\'evy flights exhibit long-range correlations and
anomalous transport \cite{Bouchaud-Georges,Levy_book}. Of course,
the completely deterministic motion of a tracer carried by a  flow
(\ref{stokes_v}) is not a random walk; the sequence of return
times is rigidly prescribed by the rotation number. Nevertheless,
as we see, already the presence of the power-law tail $T^{-3}$
appears to be sufficient for the onset of typical hallmarks of the
L\'evy flights.

In dynamics of nonintegrable Hamiltonian systems, anomalous
diffusion is ascribed to sticking of trajectories near the
Kolmogorov-Arnol'd-Moser surfaces and
Cantori~\cite{Geisel-87,Zaslavsky}. In fluid dynamics, L\'evy
flights were experimentally observed in time-dependent flows with
coherent structures (jets, vortices etc.), which serve as kind of
traps for tracer
particles~\cite{Swinney,Swinney_inbook,Venkataramani}. Our example
shows that similar anomalously fast transport can take place also
in the framework of the two-dimensional time-independent flow.
Moreover, both basic ingredients, irrational rotation number and
vanishing velocity along the whole borderline,
 are generic within the considered class of Stokes flows.
Of course, for a ``random'' rotation number $\alpha/\beta$ the peaks
of the autocorrelation would not form a regular lattice.

In conclusion, we have studied the spectral and transport
properties of a plane steady viscous flow past an array of solid
obstacles. On reducing the problem to the time-independent
integrable Hamiltonian system with 1 degree of freedom, we find
fractal Fourier spectrum, long-range correlations and anomalously
fast ``superdiffusive'' transport reminiscent of L\'evy flights.

We thank F.~H.~Busse, D.~V.~Lyubimov, A.~Pikovsky, and J.~Kurths
for fruitful discussions and H.~Hasimoto for providing the
preprint~\cite{Hasimoto-preprint}. The research of A.S. has been
enabled by the trilateral Russian-German-French program and Grant
No. PE-009-0 of CRDF, and M.Z. was supported by SFB-555.

\end{document}